\documentclass[conference]{IEEEtran}
\IEEEoverridecommandlockouts
\usepackage{ifpdf}

\ifCLASSINFOpdf
  \usepackage[pdftex]{graphicx}
\else
  \usepackage[dvips]{graphicx}
\fi
\usepackage[cmex10]{amsmath}
\usepackage{amssymb}
\usepackage{amsthm}

\usepackage{slashbox}
\usepackage{array}

\begin{document}
%
\title{Using Delta-Sigma Modulators in Visible Light OFDM Systems}

\author{\IEEEauthorblockN{Zhenhua Yu$^{1,2}$, Arthur J. Redfern$^{2}$ and G. Tong Zhou$^{1}$}
\IEEEauthorblockA{$^1$ School of Electrical and Computer Engineering,
Georgia Institute of Technology, Atlanta, GA 30332--0250, USA\\
$^2$  Texas Instruments, 12500 TI Boulevard MS 8649, Dallas, TX 75243, USA\\
Email: zhenhuayu@gatech.edu}\thanks{This paper was accepted by IEEE Wireless and Optical Communication Conference (WOCC 2014), Newark, New Jersey, May 2014}}

%


\maketitle

\begin{abstract}
Visible light communications (VLC) are motivated by the radio-frequency (RF) spectrum crunch and fast-growing solid-state lighting technology. VLC relies on white light emitting diodes (LEDs) to provide communication and illumination simultaneously. Simple two-level on-off keying (OOK) and pulse-position modulation (PPM) are supported in IEEE standard due to their compatibility with existing constant current LED drivers, but their low spectral efficiency have limited the achievable data rates of VLC. Orthogonal frequency division multiplexing (OFDM) has been applied to VLC due to its high spectral efficiency and ability to combat inter-symbol-interference (ISI). However,  VLC-OFDM inherits the disadvantage of high peak-to-average power ratio (PAPR) from RF-OFDM. Besides, the continuous magnitude of OFDM signals requires complicated mixed-signal digital-to-analog converter (DAC) and modification of LED drivers. We propose the use of delta-sigma modulators in visible light OFDM systems to convert continuous magnitude OFDM symbols into LED driver signals. The proposed system has the communication theory advantages of OFDM along with the practical analog and optical advantages of simple two level driver signals.  Simulation results are provided to illustrate the proposed system.
\end{abstract}


%
\IEEEpeerreviewmaketitle

\section{Introduction}
Visible light communication (VLC) has attracted a lot of attentions for its potential to complement conventional RF communication \cite{Brien2008, IEEEVLC, Elgala2011,Yu2013a,Jovicic2013}. VLC relies on white LEDs  which already provide illumination and are quickly becoming the dominant lighting source to transmit data. VLC is motivated by a number of benefits including, but not limited to, ``piggyback'' on existing illumination infrastructure, low-cost front-ends, more security (visible light cannot penetrate wall), no electromagnetic interference, and being safe for human. 

VLC employs simple intensity modulation (IM) and direct detection (DD) schemes, which requires the modulation signal to be real-valued and positive. Single-carrier unipolar and real-valued modulations, such as on-off keying (OOK), variable pulse-position modulation (VPPM), and pulse amplitude modulation (PAM), are adopted in VLC \cite{Elgala2011, Gancarz2013}.    
Recently, orthogonal frequency division multiplexing (OFDM) has been considered for VLC due to its ability to boost data rates and efficiently combat inter-symbol-interference (ISI) \cite{Elgala2007, Khalid2012, Grobe2013}. OFDM can easily support multiple access (OFDMA), which is essential for multi-user broadcasting. However, transmitting OFDM symbols requires the driving circuits of a white LED to support continuous magnitude inputs, and the mixed-signal digital-to-analog converter (DAC) design is complicated. In IEEE 802.15.7 standard \cite{IEEEVLC}, only two-level modulations (OOK and VPPM) are supported because their seamless compatibility with most constant-current LED drivers \cite{TILED}. Since the primary function of the VLC is providing illumination \cite{IEEEVLC, Yu2014b}, modification of driving circuits of LED will be not beneficial to the application and commercialization of VLC. Moreover, the high peak to average power ratio (PAPR) makes OFDM signals sensitive to LED nonlinearities \cite{Elgala2010, Yu2012, Yu2012b, Yu2013, Yu2014a}. 

In this work, we propose using a delta-sigma modulator \cite{Schreier2005, Redfern2013, Yu2014} to convert the OFDM signal into a two level representation and directly drive the LED. Most of the quantization noises will be pushed to the out-of-band, and the in-band subcarriers can be simply recovered at the receiver. Section \ref{SectionBackground} reviews modulation techniques in VLC and basic concept of delta-sigma digital to analog converters (DACs).  Section \ref{SectionSystem} describes the delta-sigma modulator based VLC-OFDM system and discusses the advantages.  Numerical results are shown in Section \ref{SectionResults} and conclusions are provided in Section \ref{SectionConclusions}.

\section{Background}
\label{SectionBackground}
In this section, we review visible light communications and delta-sigma digital to analog converters (DACs).

\subsection{Visible light communications}
\begin{figure}[!t]
\begin{center}
\includegraphics[width=3.5in]{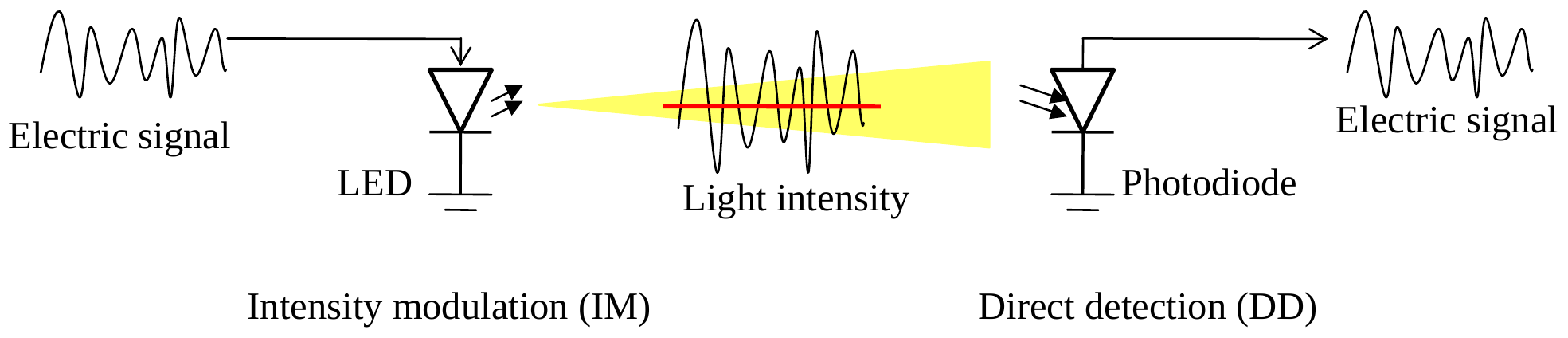}  
\caption{Intensity modulation and direct detection in VLC} 
\label{figimdd}  
\end{center}
\end{figure}

\begin{figure}[!t]
\begin{center}
\includegraphics[width=3.0in]{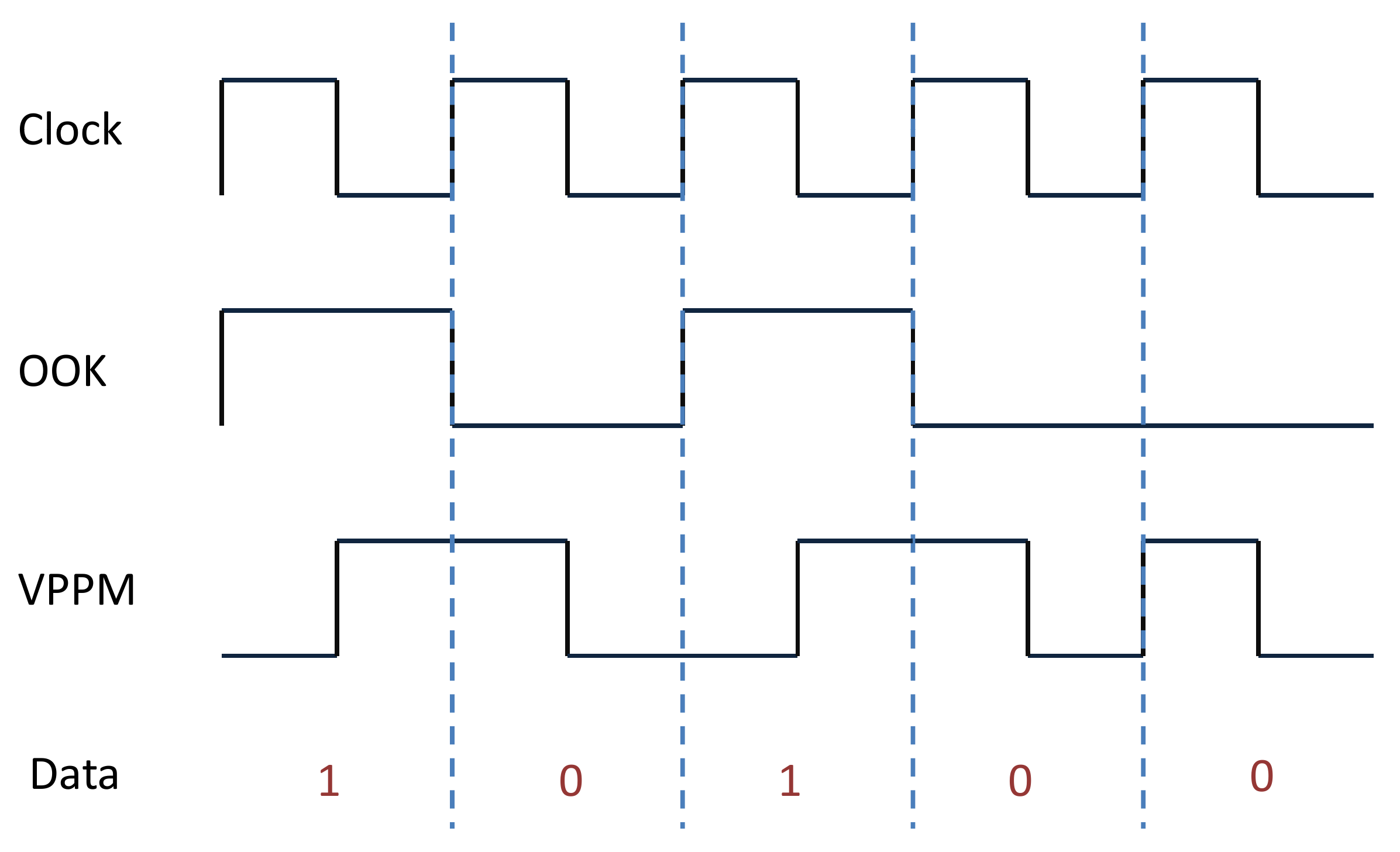}  
\caption{Examples of OOK and VPPM waveforms} 
\label{figookvppm}  
\end{center}
\end{figure}

\begin{figure*}[!t]
  \centering
  \includegraphics[width=15cm]{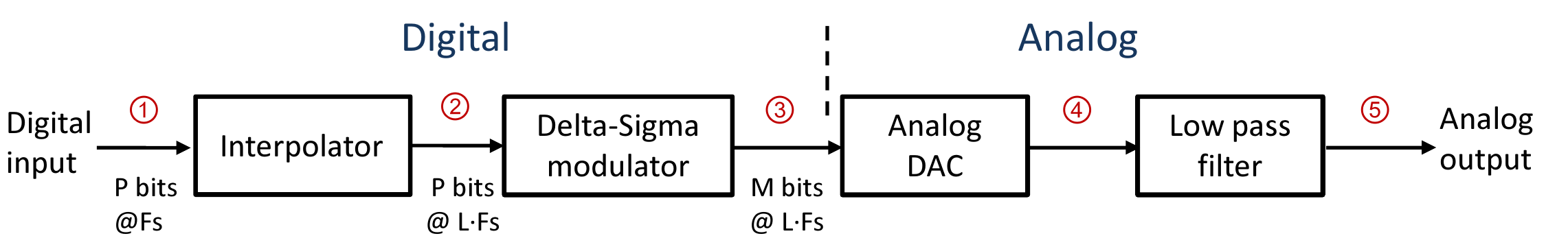}
\caption{System diagram of delta-sigma DAC.}
\label{fig_dac}
\end{figure*}

In visible light communication systems, white LED is utilized to simultaneously transmit information and illuminate. Intensity modulation (IM) is employed at the transmitter. The forward electric signal drives the LED which in turn converts the magnitude of the input electric signals into optical intensity. The human cannot perceive fast-changing variations of the light intensity, and only respond to the average light intensity. Direct detection (DD) is employed at the receiver. A photodiode (PD) transforms received optical power into the amplitude of an electrical signal. Fig. \ref{figimdd} shows the concept of intensity modulation and direct detection in VLC.

IM/DD schemes require modulation signals in VLC to be real-valued and positive. In IEEE 802.15.7 standard \cite{IEEEVLC}, on-off keying (OOK) and variable pulse-position modulation (VPPM) are supported. Fig. \ref{figookvppm} shows the waveforms of OOK and VPPM signals. OOK is the simplest modulation in VLC. OOK transmits the bit 1 by ``turning on'' the light and transmits the bit 0 by ``turning off'' the light. When transmitting the bit 0, the light is not necessarily turned off completely, but dimmed at a lower level relatively to the ``turning on'' when transmitting the bit 1. VPPM encodes the data using the position of the pulse within a time period. 0 is represented by a positive pulse at the beginning of the period followed by a lower level pulse, and 1 is represented by a lower level pulse at the beginning of the period followed by a positive pulse. 

OFDM can be applied to VLC to improve the spectral efficiency. Let $\{X_k\}_{k=-N/2}^{N/2-1}$ be the frequency domain sequence of an OFDM symbol, where $N$ is the number of subcarriers and $\Delta f$ is the subcarrier spacing. A Nyquist rate discrete time-domain
block $\mathbf{x} = [x_0, x_1, \ldots, x_{N-1}]$ is generated by
applying the inverse FFT (IFFT) operation to a frequency-domain
sequence as
\begin{eqnarray}
x_n = \text{IFFT}(X_k) =
\frac{1}{\sqrt{N}}\sum_{k=-N/2}^{N/2-1}X_k\exp\left(j2\pi\frac{ kn}{N}\right),\\\notag
n = 0, 1, \dots, N-1,
\end{eqnarray}
where $j = \sqrt{-1}$. To generate real-valued and positive baseband OFDM signal, DC biased optical OFDM (DCO-OFDM) \cite{Elgala2007} was introduced for VLC. According to the property of inverse Fourier transform, a real-valued
time-domain signal $x_n$ corresponds to a frequency-domain signal
$X_k$ that is Hermitian symmetric; i.e.,
\begin{eqnarray}
\label{eq_herm}
X_k &=& X_{-k}^*, \quad 1\leq k \leq N/2-1,
\end{eqnarray}
where $*$ denotes complex conjugate. In DCO-OFDM, the $0$th and $-N/2$th subcarrier are null; i.e., $X_0 = 0$, $X_{-N/2}=0$. Then a DC bias is added to $x_n$ to make the signal positive. One drawback of OFDM is the high peak-to-average power ratio (PAPR) \cite{Yu2013}; i.e.,
\begin{equation}
\text{PAPR}\{x_n\} = \frac{\max x_n^2}{\sum_{n=0}^{N-1}x_n^2}
\end{equation}

\subsection{Delta-Sigma DAC}

\begin{figure}[!t]
  \centering
  \includegraphics[width=8cm]{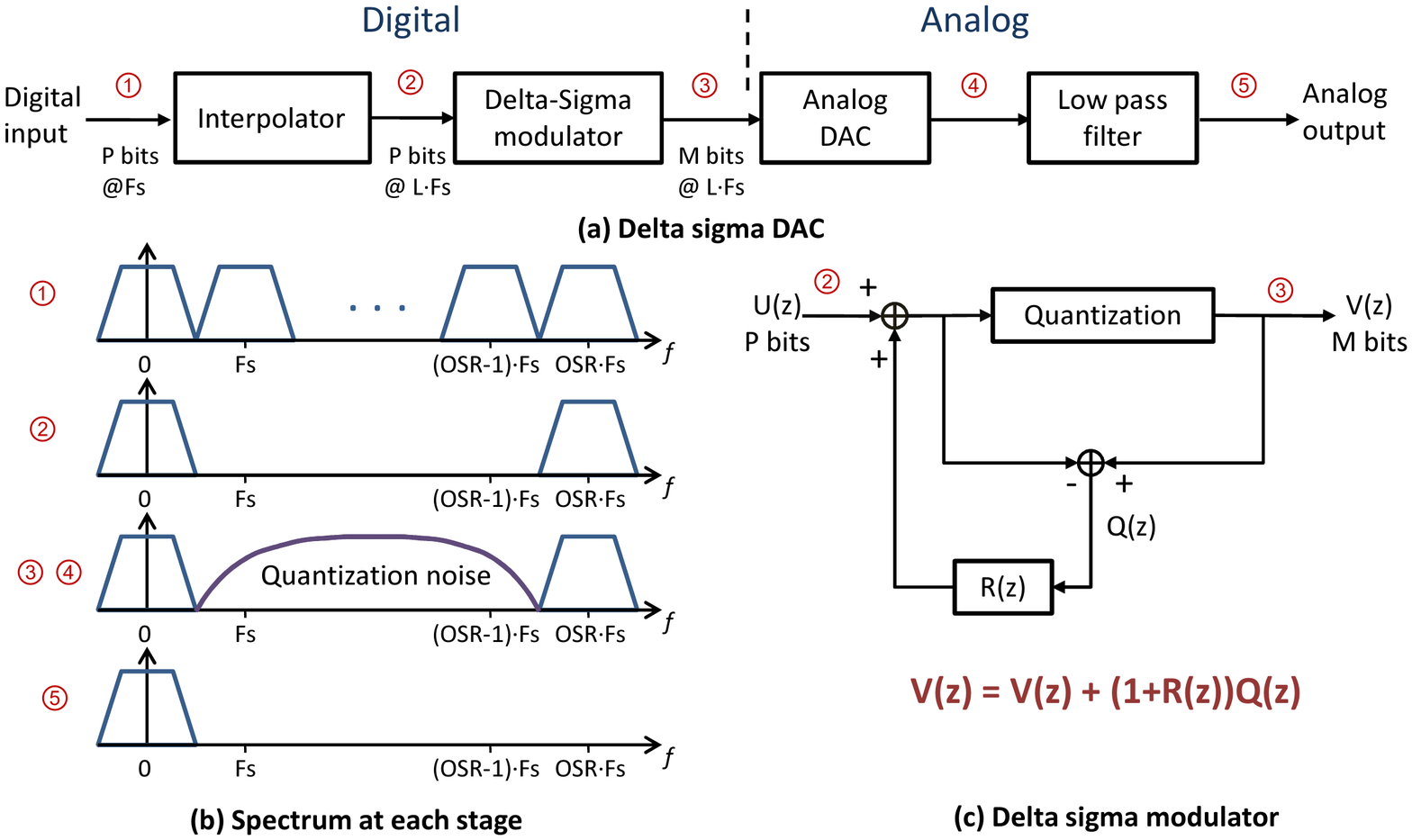}
\caption{Signal spectrum at different stages of delta sigma DAC.}
\label{fig_dacspec}
\end{figure}

\begin{figure}[!t]
\begin{center}
\includegraphics[width=3.0in]{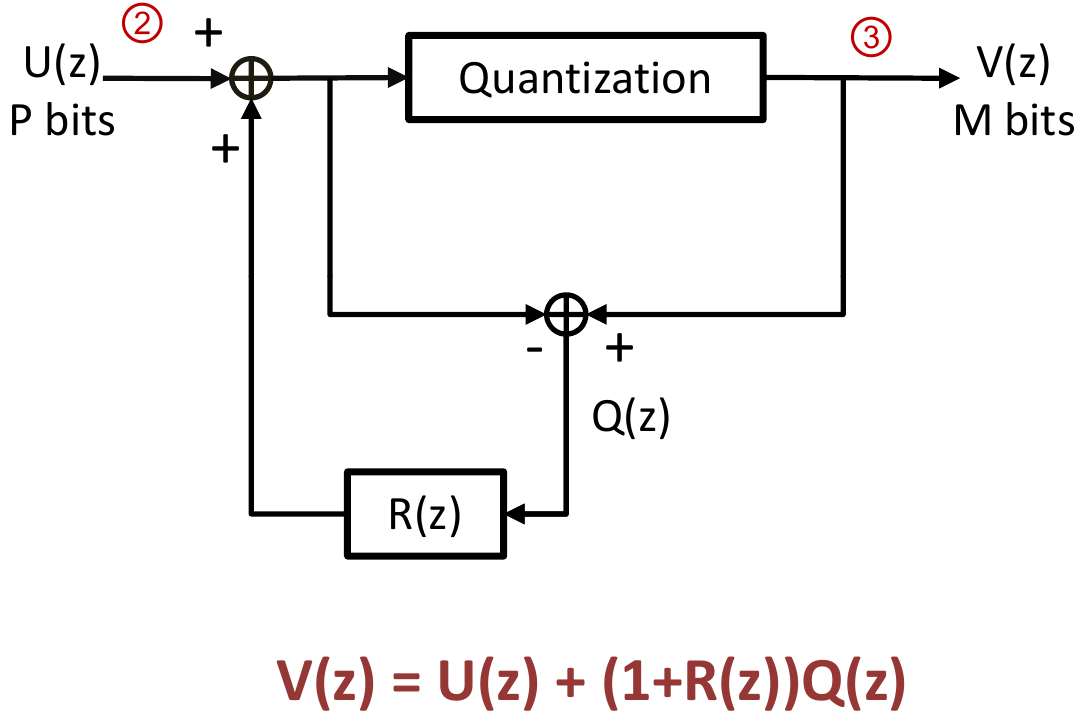}  
\caption{Delta sigma modulator.} 
\label{Figure4DeltaSigmaModulator}  
\end{center}
\end{figure}  

 \begin{figure*}[!t]
   \centering
   \includegraphics[width=16cm]{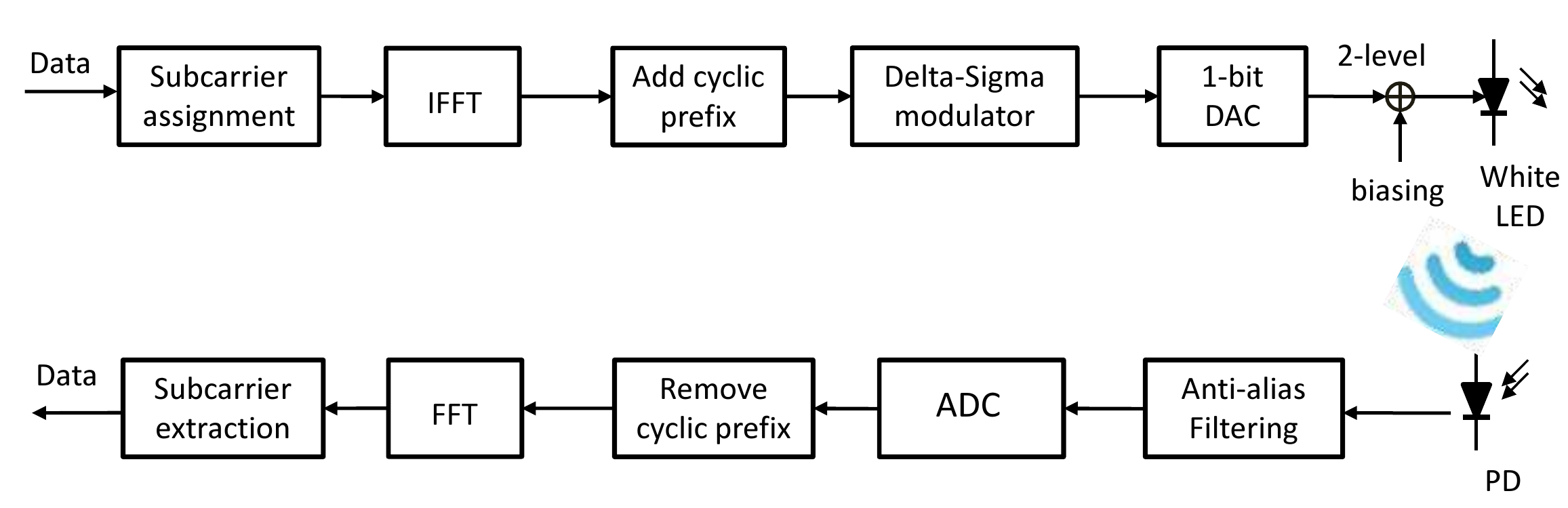}
 \caption{Visible light OFDM transmitter based on delta-sigma modulator .}
 \label{fig_vlc}
 \end{figure*}
 
Fig. \ref{fig_dac} is block diagram of a delta-sigma DAC and Fig. \ref{fig_dacspec} shows the signal spectrum at different stages. A delta-sigma DAC \cite{Schreier2005} generally comprises a interpolator, delta-sigma modulator, a mixed signal DAC core and an analog filter. The input is a digital signal sampled at rate $F_s$ with a resolution of $P$ bits. The interpolator upsamples the signal to sampling rate $L\cdot F_S$, where $L$ is the over sampling ratio (OSR), and suppress the spectral replicas centered at $F_s,\,2F_s\,\dots,\,(L-1)F_s$. 

The role of a delta-sigma modulator is to convert a digital signal with $2^P$ levels into a digital signal with $2^M$ levels, where $M < P$,  while maintaining a high in-band signal-to-noise power ratio. Converting to a signal with only a few levels makes the analog DAC more tolerant to component mismatch and nonlinearities. 
Fig. \ref{Figure4DeltaSigmaModulator} shows the block diagram of a delta-sigma modulator.  The operation of delta-sigma modulator can be expressed in the z domain as
\begin{equation}
\label{inoutdsmeq}
V(z) = U(z) + (1+R(z))Q(z)
\end{equation} 
where $U(Z)$, $V(Z)$, and $Q(Z)$ denote the $z$ transform of the input, output and quantization error of the delta-sigma modulator, respectively. $1+R(z)$ is noise transfer function (NTF).


The delta-sigma modulator pushes quantization noise out of the signal band through the appropriate design of the NTF. A mixed signal DAC reproduces the $2^M$ level digital signal at its input and an analog lowpass filter removes most of the out-of-band noise power and creates a $P$ bit resolution analog signal.

\section{Visible light OFDM transmission based on a delta-sigma modulator}
\label{SectionSystem}

\begin{figure}[!t]
\begin{center}
\includegraphics[width=3.2in]{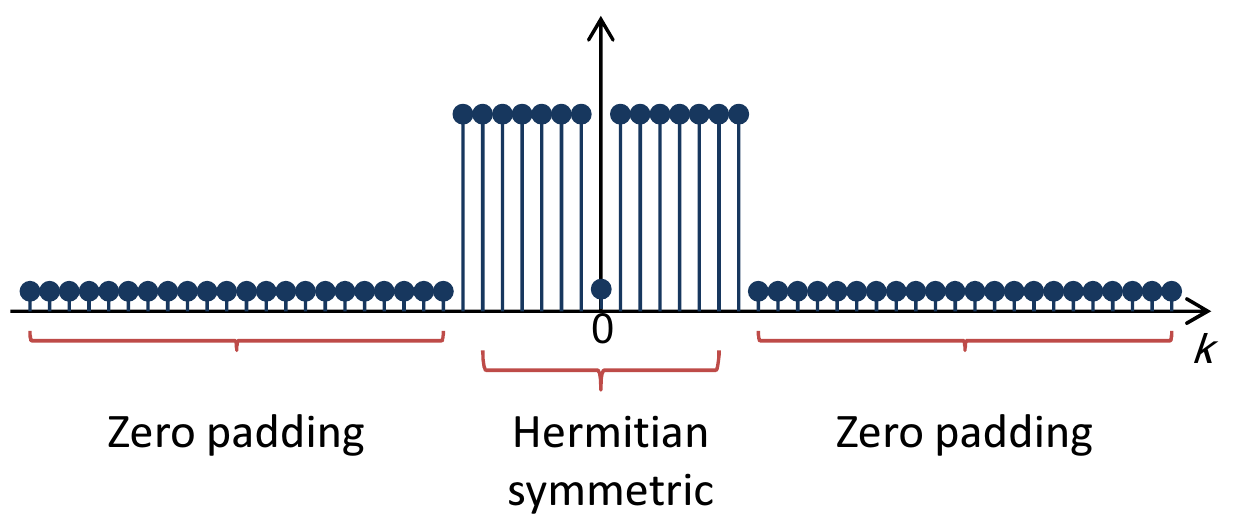}  
\caption{Subcarrier assignment for oversampled VLC-OFDM signal.} 
\label{Figure7subcarrier}  
\end{center}
\end{figure}

\begin{figure}[!t]
\begin{center}
\includegraphics[width=3.1in]{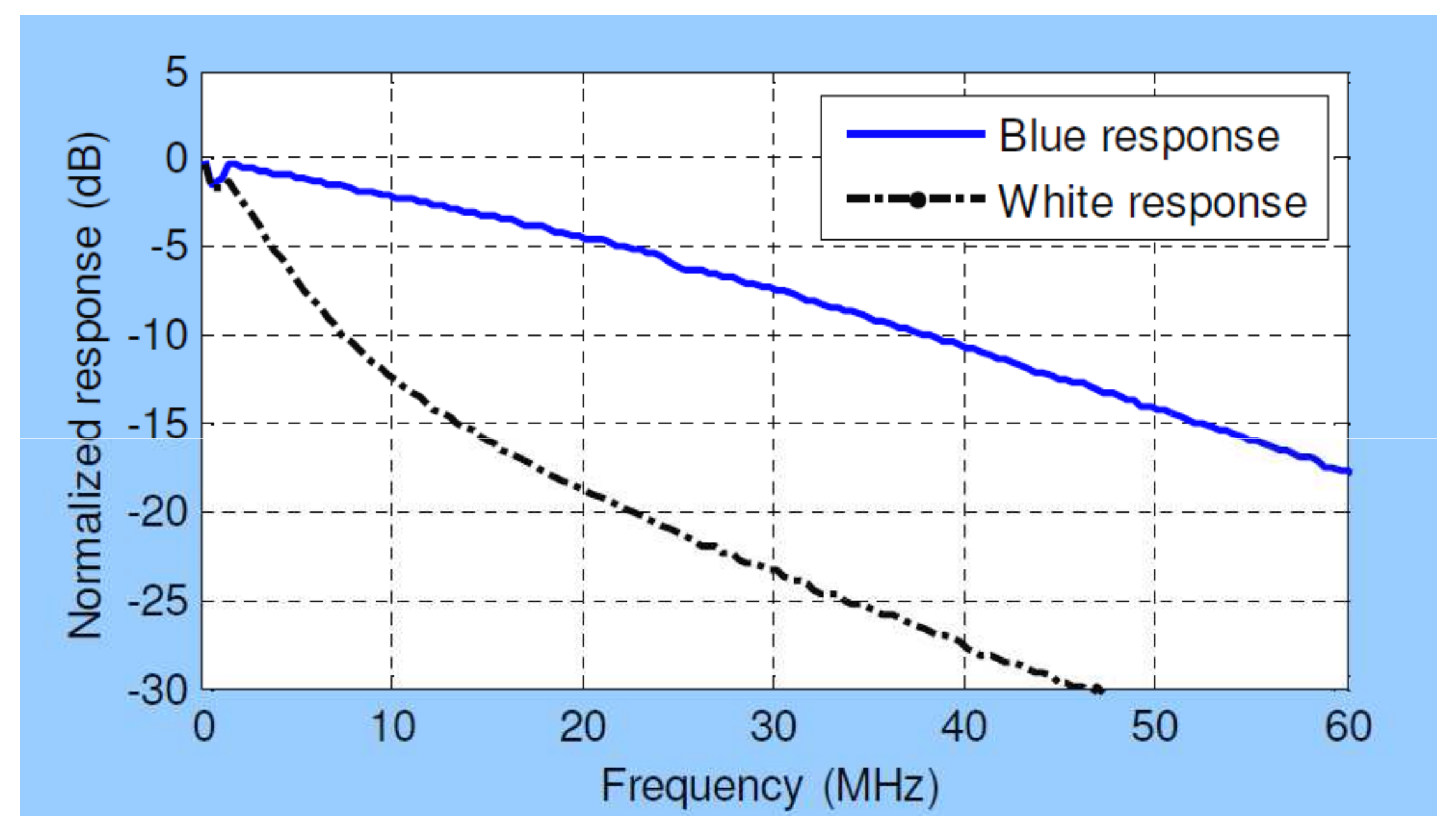}  
\caption{frequency response of emitted white light and the blue part of a typical white light (Luxeon STAR) LED \cite{Minh2009}.} 
\label{Figure8ledfreq}  
\end{center}
\end{figure}  
 \begin{figure*}[!t]
   \centering
   \includegraphics[width=13cm]{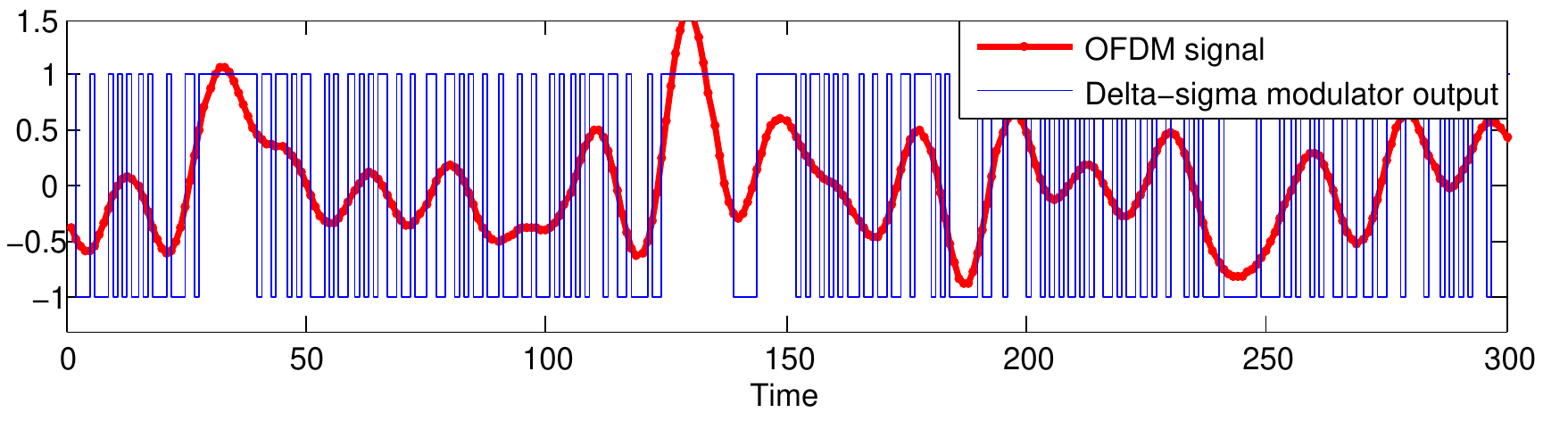}
 \caption{Input and output sequence of delta-sigma modulator (L = 8).}
 \label{fig_time}
 \end{figure*}
Fig. \ref{fig_vlc} is a block diagram of visible light OFDM transmission that uses a delta-sigma modulator. 
In a delta-sigma DAC, interpolation is used for generating oversampled signals at the input of the delta-sigma modulator. In our proposed framework, we obtain the oversampled OFDM signals by zero padding in the frequency domain. Define the in-band indices to be the set $\mathcal{I} = [-N/2, N/2-1]$, and the out-of-band indices to be the set $\mathcal{O} = [-NL/2,-N/2-1] \cup [N/2,LN/2-1]$. The zero padded version of $X_k$ can be expressed as
\begin{equation}
X_k^{(L)}=
\begin{cases}
X_k, & k \in \mathcal{I} \\
0, &  k \in \mathcal{O}
\end{cases}.
\end{equation}
Fig. \ref{Figure7subcarrier} shows the subcarrier assignment for oversampled VLC-OFDM signal.
An $LN$ length IFFT is used to convert the frequency domain sequence $\{X_k^{(L)}\}_{k=-LN/2}^{LN/2-1}$ into a $L$ times oversampled time-domain sequence $\{x_n^{(L)}\}_{n=0}^{LN-1}$.
 
The delta-sigma modulator is then used to convert the continuous magnitude time domain OFDM digital samples $\{x_n^{(L)}\}_{n=0}^{LN-1}$ into one bit signals $\{y_n^{(L)}\}_{n=0}^{LN-1}$, where $y_n^{(L)} \in \{-1,1\}$. According to Eq. (\ref{inoutdsmeq}) , the data on the frequency-domain subcarriers of $y_n^{(L)}$ can be expressed as 
\begin{eqnarray}
Y_k^{(L)} &=& X_k^{(L)} + \text{NTF}_kQ_k\\\notag
&=&
\begin{cases}
X_k + \text{NTF}_kQ_k, & k \in \mathcal{I} \\
\text{NTF}_kQ_k, &  k \in \mathcal{O}
\end{cases},
\end{eqnarray}
where $\text{NTF}_k$ denotes the noise transfer function on the $k$th subcarrier and $Q_k$ denotes the quantization noise on the $k$th subcarrier, respectively. Since NTF is actually a high-pass filter, very few distortions fall on the in-band subcarriers. We use error vector magnitude (EVM) \cite{Yu2012,IEEE80216} to quantify the in-band distortions
\begin{eqnarray}
\text{EVM} &=& \sqrt{\frac{\sum_{k=1}^{N/2-1}|Y_k^{(L)} - X_k^{(L)}|^2}{\sum_{k=1}^{N/2-1}|X_k^{(L)}|^2}}\\\notag
&=& \sqrt{\frac{\sum_{k=1}^{N/2-1}|\text{NTF}_kQ_k|^2}{\sum_{k=1}^{N/2-1}|X_k|^2}}.
\end{eqnarray}

The proposed DAC has a single bit digital input and a two level analog output. To ensure that the input of the LED is positive, a bias is added to the two level analog signal, which only affects the DC component. Note that this bias could be built into the DAC.
 
We directly use the two level DAC output signal (after biasing) that contains out-of-band quantization noise to  drive the LED for two reasons. First, most of the existing driving circuits of white LEDs used for the purposes of illumination are not compatible with continuous amplitude driving voltages. However, a two level input signal, like a pulse width modulation (PWM) or OOK signal, is widely accepted \cite{IEEEVLC}. Second, out-of-band inference is not a concern in VLC because the LED acts as a low pass filter (similar to a speaker in a class D amplifier audio system \cite{Dapkus2000}) and no out-of-band  information is transmitted simultaneously.

As an example, the most popular white LED, which uses a blue emitter in combination with a phosphor that emits yellow light, has limited bandwidth. The frequency response of emitted white light and the blue part of a typical white light (Luxeon STAR) LED is shown in Fig. \ref{Figure8ledfreq} \cite{Minh2009}. The bandwidth of the white light response is only 2.5 MHz. Any remaining out-of-band noise can be removed or ignored at the receiver, either with a low pass anti-aliasing filter or by ignoring out-of-band subcarriers after the FFT. 

The delta-sigma modulator based transmitter also has some advantages over a conventional visible light OFDM transmitter. First, the structure of a 1 bit mixed signal DAC is very simple and its linearity is theoretically perfect. Second, the drawback of the high PAPR of OFDM signal is avoided. Third, it is unnecessary to design a customized driving circuit for the LED to support continuous or multiple amplitude level input signals.  

\section{Numerical results}
\label{SectionResults}
\begin{figure}[!t]
\centering
\includegraphics[width=8cm]{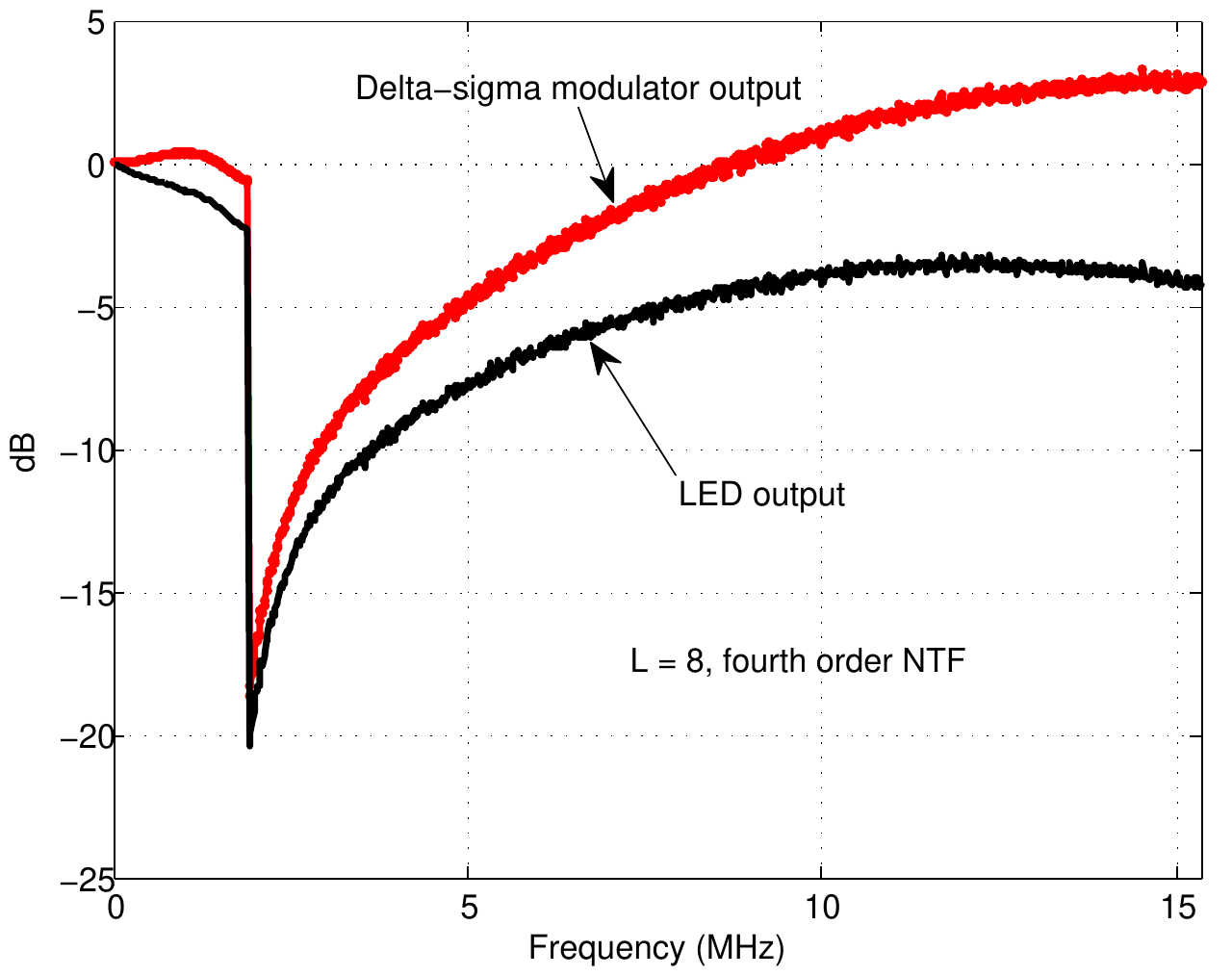}
\caption{PSD of input and output of LED.}
\label{fig_psd}
\end{figure}

\begin{figure}[!t]
\centering
\includegraphics[width=8cm]{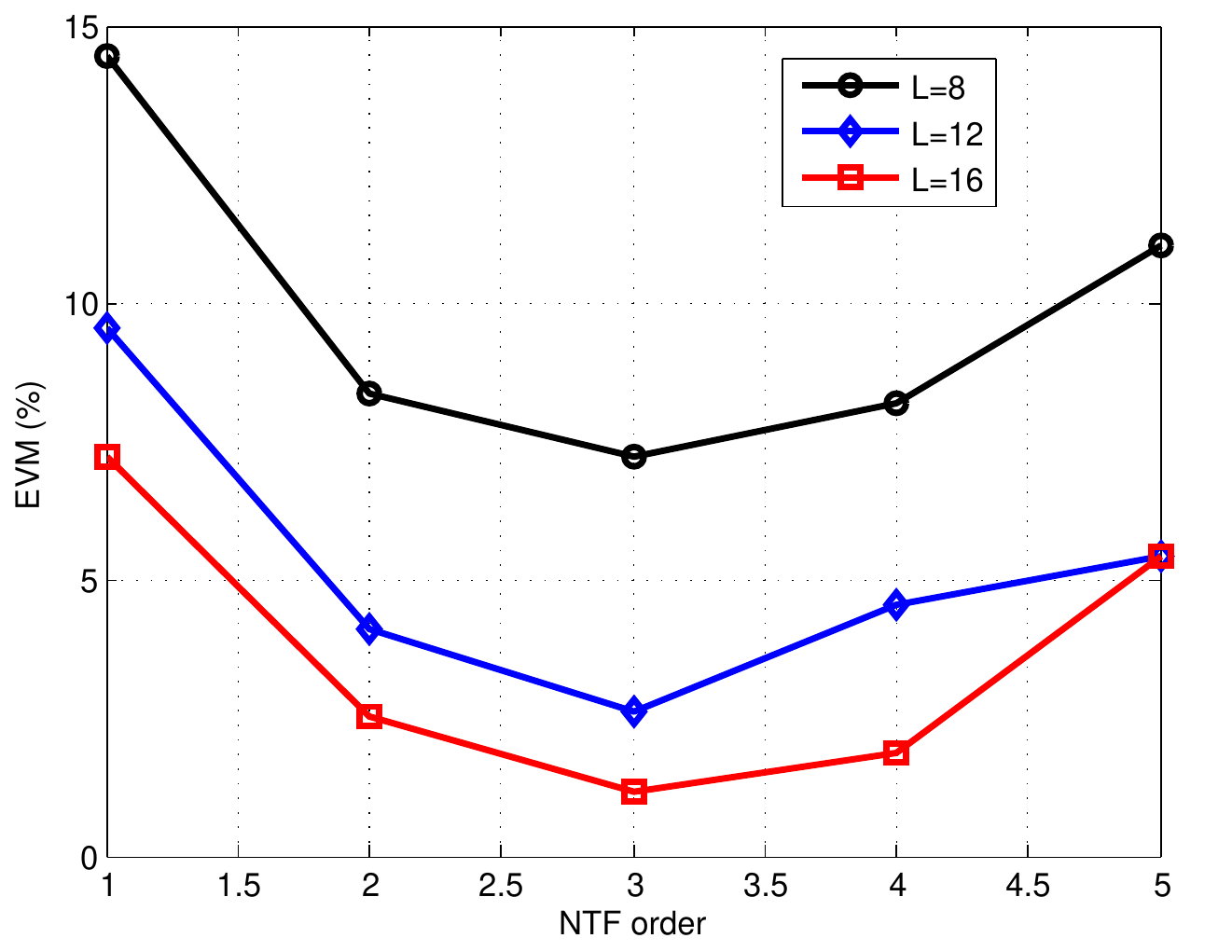}
\caption{Error vector magnitude of transmitted signals.}
\label{fig_evm}
\end{figure}

To illustrate the proposed system, we choose $N = 256$ subcarriers with a spacing of $\Delta f  = 15$ kHz, and 4-QAM modulation. The bandwidth of the OFDM signal, 1.875 MHz, is within the 3 dB bandwidth of the white LED from \cite{Minh2009}. The NTF of the delta-sigma modulator was obtained from the Matlab toolbox associated with \cite{Schreier2005}. 
Fig. \ref{fig_time} plots a section of the input and output of the delta-sigma modulator and Fig. \ref{fig_psd} plots the power spectral density (PSD) of the input and output of the LED with a fourth order NTF and $L=8$. Note that most of the noise falls out of the signal band. 
Fig. \ref{fig_evm} plots the EVM of transmitted signals with various OSRs and orders of NTF. By increasing the OSR, less in-band noise is observed. However, increasing the order of NTF does not always guarantee better performance because the stability of the loop filter in a delta-sigma modulator may become a problem. Different loop filter designs can be created to address this issue. We can observe that  if we choose OSR greater than 12, EVM can be as low as 2\%, which is much less than the maximum allowed EVM thresholds in most IEEE standards \cite{IEEE80216, Petrick2005}.

%


\section{Conclusions}
\label{SectionConclusions}
We used a delta-sigma modulator to convert a continuous magnitude OFDM digital signal into a two level analog signal that can directly serve as the input of a LED. This scheme eases the design of the mixed signal DAC and driving circuits, as well as avoids nonlinear distortion due to a high PAPR. 

\section*{Acknowledgment}

This research was supported in part by the Texas Instruments Leadership University Program.







%
\end{document}